# Spectral radiative analysis of bio-inspired H₂ production in a benchmark photoreactor: a first investigation using spatial photonic balance

Caroline SUPPLIS[1], Fabrice GROS[1], Ghiles DAHI[1], Jérémi DAUCHET[1], Matthieu ROUDET[1], Frédéric GLOAGUEN[2], Jean-François CORNET[1]

[1] Université Clermont Auvergne, CNRS, Sigma Clermont, UMR 6602, Institut Pascal, F-63000 Clermont-Ferrand, France

[2] UMR 6521, CNRS, Université de Bretagne Occidentale, CS 93837, 29238, Brest, France

**Corresponding author :**

Fabrice GROS

Université Clermont Auvergne, CNRS, Sigma Clermont, UMR 6602, Institut Pascal BP 1044, Clermont-Ferrand 63000, France. Fax: +33473407829.

E-mail: fabrice.gros@sigma-clermont.fr

**Highlights**

A bioinspired photocatalytic system producing $H_2$ in a photoreactor was evaluated.

A spectral radiative analysis of the system was carried out.

Evidence was obtained that linear kinetic coupling with light absorption rate occurs.

The bio-inspired catalyst reveals strong capabilities from an engineering analysis.

**Abstract**

The iron–thiolate complex [Fe$_2$(μ-bdt)CO$_6$] (bdt =1,2-benzenedithiolate) is a simplified model of the active site of diiron hydrogenase enzymes. Here, we describe the implementation of this noble-metal-free catalyst in aqueous solutions using eosin Y as photosensitizer and triethylamine as an electron donor in an experimental bench specially designed for the study of H$_2$ photoproduction. The bench is composed of an adjustable visible light source, a fully equipped flat torus photoreactor and analytical devices. Rates of H$_2$production under varied experimental conditions were obtained from an accurate measurement of pressure increase. A spectral radiative analysis involving blue photons of the source primarily absorbed has been carried out. Results have proven the rate of H$_2$ production is proportional to the mean volumetric rate of radiant light absorbed demonstrating a linear thermokinetic coupling. The bio-inspired catalyst has proven non-limiting and reveals interesting capabilities for future large scale H$_2$ production.

**Keywords**

Hydrogen, spectral radiative transfer analysis, photoreactor, overall quantum yield, specific rate, bioinspired catalyst.

**Introduction**

The growth of global energy demand due to increasing population and rising standards of living create an acute energetic and environmental issue [1]. At this moment, Humanity is largely dependent on fossil fuels that provide up to 85% of the energy demand [2]. These resources are not inexhaustible and their utilizationgenerates$CO_2$emission. Utilization of hydrogen ($H_2$) seems to be one of the most promising spare solutions provided that this fuel is produced with environmentally friendly technologies and from renewable resources, such as from water and solar energy for instance. This Grail however is far to be trivial and is today facing both fundamental and engineering challenges in order to gain two or three orders of magnitude for kinetic and energetic performances at industrial scale.

First, regarding the point of view of the catalysis, a widespread energy transformation process based on water and photons is the natural photosynthesis, in which water is split into oxygen and protons at a photosystem. The protons can then be reduced into $H_2$at specific enzymes called hydrogenases present in cyanobacteria and algae. Although photosystem and hydrogenase enzymes are very efficient catalysts *in vivo*, they rely on a delicate protein framework and are sensitive to subtle changes in their environment, making them largely unsuitable for use under a wide variety of conditions[3].

Hence, a large field of research called artificial photosynthesis is dedicated to the development of cheap catalysts which mimic the function of natural enzymes. Currently, research efforts are mainly focused on the independent development of photocatalytic systems for the two half-reactions, i.e. the combination of a photosensitizer (P) with a suitable catalyst (Cat) and with a sacrificial electron acceptor (A) or donor (D) for either the oxidation of water or the reduction of protons(Fig. 1)[3–5].

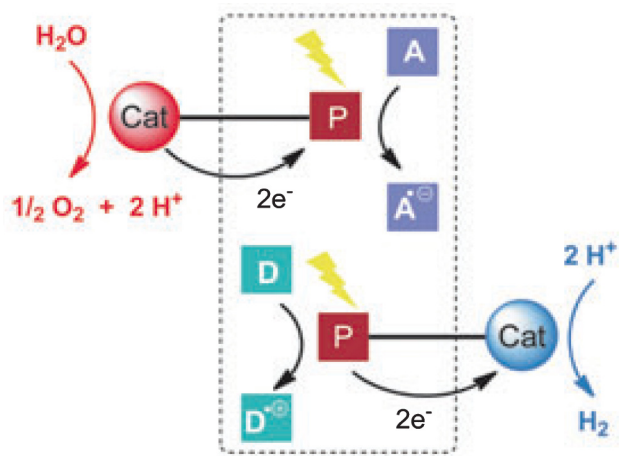

Fig. 1 – Operating principle of chemical systems for artificial photosynthesis: water oxidation (top) and proton reduction(bottom) [6]

Hereafter, we will focus on photocatalytic systems for $H_2$ production, of which numerous examples have been described in the literature[7–14]. Over the last decade, molecular catalysts formed from earth-abundant materials have been revisited with the hope that they will in the future compete favorably with noble-metal-based catalysts. Significant advances in this area have been achieved with iron–thiolate complexes of the type $[Fe_2(\mu\text{-SRS})(CO)_6]$ (R = organic group) which are simplified models of the active site of diiron hydrogenase enzymes[15].

Second, from the engineering point of view, and considering that photo-reactive processes are controlled at different scales by radiative transfer (or photon transport), the understanding and optimization of the photon absorption step is also essential to the efficient production of $H_2$ in a photocatalytic system. It is therefore necessary to gain further insights into this physical phenomenon, which can be quantified properly by determining the local and spatial volumetric rates of radiant energy absorbed [16–20].This quantification also serves as a basis for the formulation of coupling laws between chemical reaction and photon absorption rates. Describing the rate of $H_2$ production using the mean volumetric rate of energy absorbed could hence improve the comprehension of the behavior of the photocatalytic system and facilitate the scaling-up.

Herein, we present a radiative analysis of $H_2$ production by a bioinspired and noble-metal-free photocatalytic system using a sacrificial electron donor. The first part of the article describes the material and experimental methods used in this work. The second part deals with a theoretical radiative description of the photocatalytic system. The major outcomes of the work in terms of kinetic and energetic aspects are presented in the last part of the paper.

## 1. Photo-reactive system description, Material and methods

### 1.1. Photocatalytic system

The photocatalytic system for hydrogen production implemented in our laboratory scale photoreactor was prepared from an aqueous Sodium Dodecyl Sulfate (SDS) solution of aniron-thiolate complex as a proton reduction catalyst ($[Fe_2(\mu\text{-bdt})(CO)_6]$, Fig. 2, left), eosin Y as photosensitizer ($EY^{2-}$, Fig. 2, right), and triethylamine ($Et_3N$) as a sacrificial electron donor [21]. Note that the catalyst is not soluble in water, hence we need to add SDS to prepare aqueous micellar solutions in which the concentration of catalyst does not exceed 0.1 mM [21,22].

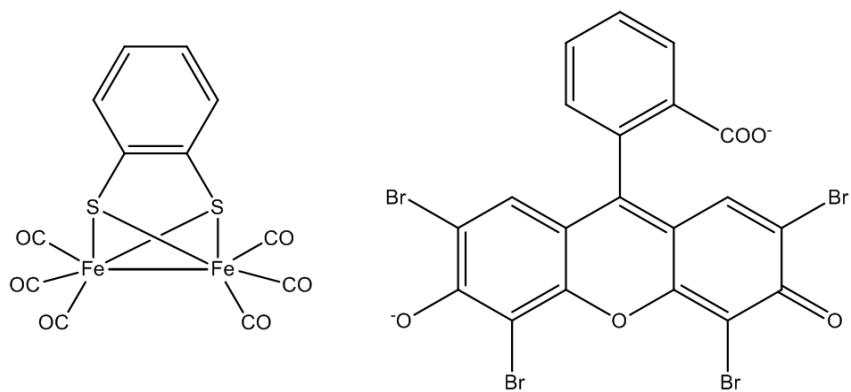

Fig. 2 – Chemical structures of the iron-thiolate complex [Fe$_2$(µ-bdt)CO$_6$] (left) and eosin Y (right) [21]

The iron-thiolate complex [Fe$_2$(µ-bdt)(CO)$_6$] was synthesized as previously described [22]. HCl (37% in weight), SDS, Et$_3$N, MeOH, and EY$^{2-}$ were purchased from Sigma Aldrich and used as received. Deionized water (Milli-Q plus) was employed to prepare the solutions after elimination of dissolved gases using an ultrasonic bath (Fisher Scientific FB 15061) under vacuum for one hour.

Two stock solutions at a concentration of 6 mM of EY$^{2-}$ and [Fe$_2$(µ-bdt)(CO)$_6$] both in MeOH were prepared and placed during 30 minutes in the ultrasonic bath to entirely dissolve the compounds. A 0.5 M aqueous solution of SDS was prepared. A mixture of Et$_3$N (10% vol.) in water was acidified at pH 10.5 with concentrated HCl. The photoreactor was then filled with an intermediate solution containing [Fe$_2$(µ-bdt)(CO)$_6$], SDS and Et$_3$N and with the solution containing EY$^{2-}$. The respective volumes of each solution were calculated to achieve the desired [Fe$_2$(bdt)(µ-CO)$_6$] and [EY$^{2-}$] final concentration (see Fig. 9). The solution in the photoreactor was finally purged for 1 hour under an argon flow to remove dissolved oxygen.

Such photoreactive system can operate in water with a mechanism proposed as follow (Fig. 3):

a) When $EY^{2-}$ absorbs photon energy, electrons are excited to higher electronic level before rapidly relaxing in the first electronic state (the singlet $^*EY^{2-}$).

b) In the most favorable case for hydrogen production, electrons undergo a spin conversion by intersystem-crossing (ISC) into triplet state ($^{3*}EY^{2-}$), a stable long lived species[23,24].

c) Based on this triplet state, since potential for the couple $EY^-/^{3*}EY^{2-}$ (-0.87 V) is more negative than the reduction potential of catalyst complex $[Fe^IFe^I]/[Fe^0Fe^I]$ ($E_{1/2}$=-0,74 V vs SHE at pH 7) [21,25], an electron is transferred to the diiron catalyst (Fig. 3).

d) To close eosin Y cycle, $EY^-$ produced by the oxydo-reduction reaction between $^{3*}EY^{2-}$ and catalyst, is regenerated in $EY^{2-}$ by an electron transfer from $Et_3N$.

e) A succession of oxidation-reduction and protonation with the catalyst, extensively described and analyzed in reference [25,26], arises to produce hydrogen.

f)

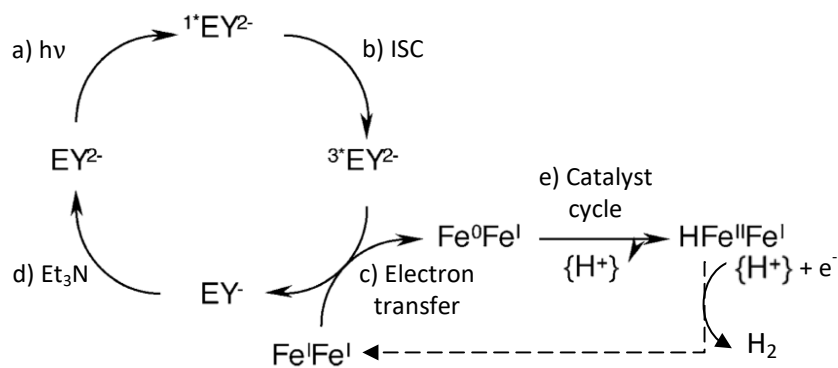

Fig. 3 – Mechanism for the formation of the iron hybrid intermediate proposed in[21,25,26].

Undesirable phenomena for hydrogen production are the following:

- Either $EY^{2-}$ electrons in the singlet state ($^*EY^{2-}$) can go down to the ground state by fluorescence emission (a spontaneous light emission) or it can relax by non-radiative emission, thermalization or quenching. This last phenomenon is caused by collisions between $^*EY^{2-}$ with other molecules (quenchers) such as $Et_3N$, $H_2O$, or eosin Y itself for instance. Quenching results in decreasing singulet state concentration and fluorescence intensity. Stern-Volmer equation studies on other diiron photocatalyst systems have established effects of quencher concentration such as dye [9,27], catalytic complex [9,10] and amine electron donor [10,28–30] on fluorescence intensity.

- Assuming $^*EY^{2-}$ is converted into $^{3*}EY^{2-}$ by ISC, relaxing phenomena to the ground state by phosphorescence doesn't occur at 25°C but quenching still exists resulting in the decreasing of the triplet state concentration[28,31–35].

### 1.2. Photoreactor

The laboratory scale photoreactor has been extensively described and characterized elsewhere[36].To sum up, the photoreactorof 190 mL ± 3 mL used in this study is a flat torus and gas tight reactor with two translucent faces made of glasses at the front and at the rear.This configuration allowed us to measure the photon flux densities exiting at the rear and to acquire spectra with the devices presented in section 1.3.

Two pneumatic valves (Carten CMDA250) at the inlet and outlet of the reactor allowed closing and opening the reactor for inerting or purging. Lateral inlets/outlets permitted the loading/evacuation of the reactor content. The fluid was stirred in the annular space ata high rotation speed of 1000 rpm by the mean of a helical impeller connected to a micromotor 24V/DC (Minisprint Magnetic Stirrer made by Premex Reactor ag) in order to reduce the characteristic time constant for the $H_2$ mass transfer phenomena. In these conditions, the photoreactor is a perfectly mixed reactor[36].The temperature of the reactor was controlled at 25.0 ± 0.1 °C by water circulation coming from a thermostatic bath (Lauda eco RE 415) connected to a temperature RTD Pt100 sensor (TC direct) and immersed in the reactor. A pressure sensor (Keller PA 33X) located in the headspace of the reactor and connected to a converter (K107)was used for the accurate detection of gas production *via* pressure increase. Keller software (Read 30) allowed for monitoring and recording of the pressure variation. A micro gas chromatograph (Agilent 3000A Micro GC) equipped with a molecular sieve(MS-5A 14 m) and a thermal conductivity detector (TCD) with argon as gas carrier was positioned at the reactor gas outlet to analyze online the composition of the reactor gaseous phase. A simplified drawing and photograph of the photoreactor equipped with its devices are represented onFig. 4; more details could be found in reference[36].

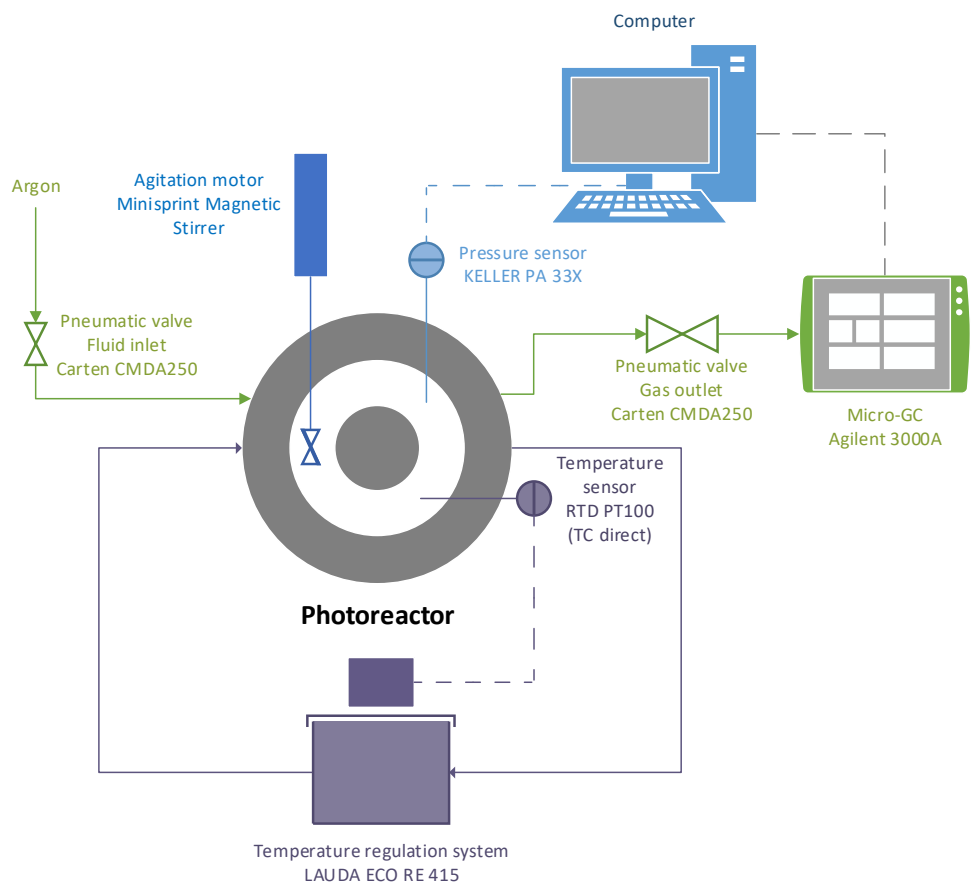 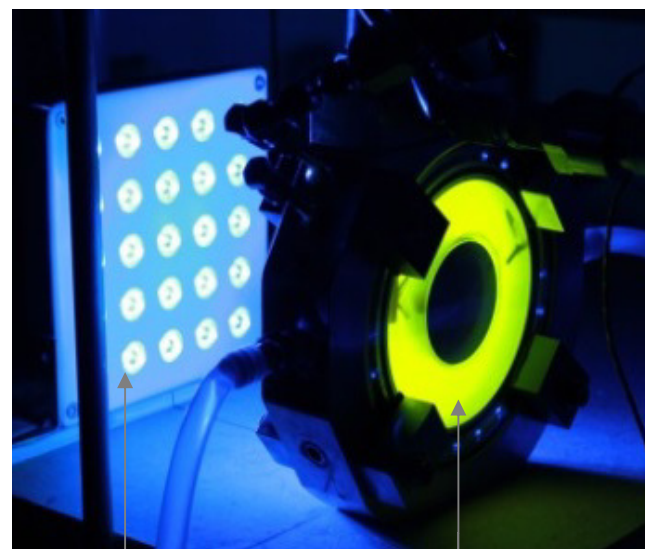

a)             b)

Fig. 4 – a) Simplified representation of the experimental set up -b) photograph of the photoreactor containing the reaction medium studied in this publication illuminated by the blue LED panel.

### 1.3. Light source characterisation and spectral light flux density measurements

The photoreactor was illuminated by a 25 LED panel (Royal blue D42180, Seoul semiconductor), positioned at a distance of 15 cm from the reactor, providing a normal quasi-collimated blue light with an emission maximum at 457 nm wavelength (Fig. 5). In such a configuration(normal collimated incident light on a purely absorbing non scattering medium) and taking into account the photoreactor geometry, the radiative transfer theory can be properly approximated as a one-dimensional problem.

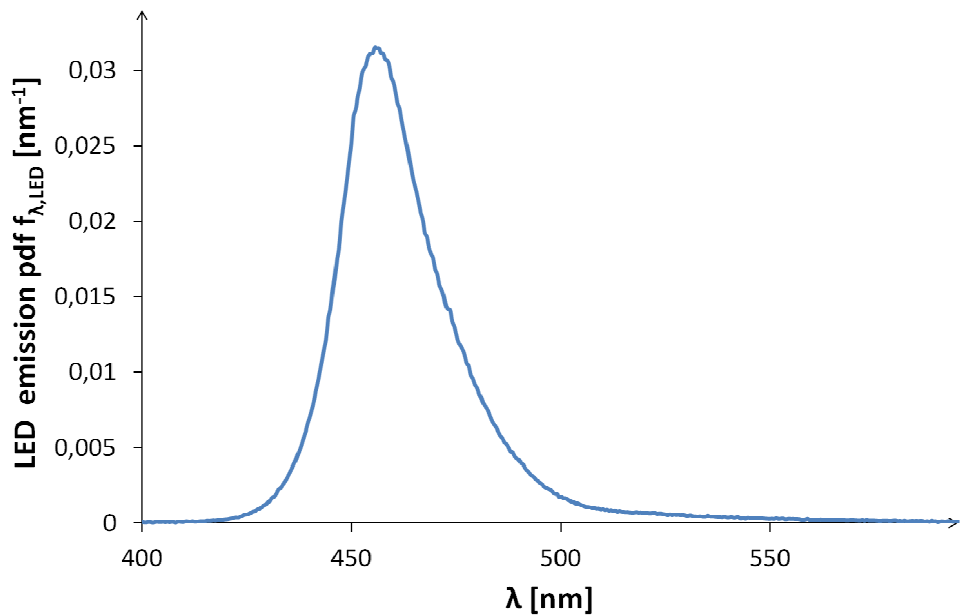

Fig. 5 – LED emission spectrum, i.e. wavelength probability density function (pdf) $f_{\lambda,LED}$.

Photon flux density was accurately controlled *via* an USB DMX controller and the Easy standalone software (Nicolaudie), including 256 different setting positions to modify the electric power supplied to the LED and so the hemispherical photon flux density $q_\cap$ entering the photoreactor (determined by sensor mapping and actinometry as explained in[36,37]).

To carry out a radiative analysis of the photocatalytic system in the photoreactor, we measured the mean photon flux density exiting of the reactor, $q_{out}$, estimated at the rear glass window with a LI-COR quantum sensor (LI-190) connected to a portable meter LI-CORQuantum/Radiometer/Photometer LI-189. This mean photon flux density at the rear of the photoreactor, $q_{out}$, was obtained by measuring at 10 different positions the photon flux density at the rear. However, this device counts only the total hemispherical photon flux density and was not able to determine the photon spectral distribution. Thus our experimental bench was complemented with a spectrometer (Ocean Optics USB 2000+) equipped with an optical fibre (QP 400-2-SR) and a cosine corrector (Ocean Optics CC-3) coupled to the fibre. The spectrometer was linked with USB connection to a computer running the Oceanview software. Spectral distribution of the photons at the rear of the reactor was thus obtained. An example can be seen in Fig. 6.

Finally, in order to determine the molar absorption coefficients, Ea, of each absorbing species present in the photoreactor (which will be useful for radiative transfer analysis)absorption spectra were carried out with a spectrophotometer UV-160A Shimadzu.

### 1.4. Hydrogen reaction rate measurement

Hydrogen was produced in the gas tight reactor under irradiation; this was indicated by the increase of the pressure in the headspace. To analyze this phenomenon, a complete mass balance on $H_2$ was developed taking into account a transient and a linear regimes[36,38]. During the linear regime, the reaction rate $\langle r_{H_2} \rangle$ can be calculated from the linear variation of pressure P with time t (Equation (1)):

$$\langle r_{H_2} \rangle = \left( \frac{V_G}{V_L RT} + \frac{1}{H_{H_2}(T)} \right) \frac{dP}{dt} \qquad (1)$$

The photoreactor is filled with a liquid volume $V_L$ of 160 mL, consequently the headspace gas volume was $V_G$ equal to 30 mL. The Henry constant $H_{H_2}(T)$ was estimated to a value of 130 770 $m^3.Pa.mol^{-1}$ at 25°C in a mixture of $Et_3N$ (10% vol.) in water[39–42].

## 2. Description of the radiative approach for kinetic and thermodynamic analysis

### 2.1 Radiative coupling model at spatial scale

As previously explained in introduction, because of the existence of a field in volumetric rate of radiant energy absorbed $\mathcal{A}$ inside any photoreactor, a physical or engineering treatment of photo-reactive systems relies on a local thermokinetic coupling involving at least an overall quantum yield, $\Phi$, with the general form[18,20,43] :

$$r_{H_2}(\mathbf{r}) = \Phi \mathcal{A}(\mathbf{r}) \qquad (2)$$

Nevertheless, if this quantum yield does not depend in any case of the radiation field $\mathcal{A}$, averaging these two rates at the spatial scale of the photoreactor (the scale of physical observable and measured quantities) is straightforward and leads to:

$$\langle r_{H_2} \rangle = \Phi \langle \mathcal{A} \rangle \qquad (3)$$

where we have introduced for convenience the bracket notation $\langle \ \rangle = \frac{1}{V}\iiint_V \, dV$. This means that, in these conditions, a plot of $\langle r_{H_2} \rangle$ versus $\langle \mathcal{A} \rangle$ (obtained for different incident light fluxes or dye concentrations) is a straight line, the slope of which being the overall quantum yield. At the opposite, if the thermokinetic coupling is non-linear, a mechanistic or phenomenological analysis of the coupling is required in order to formulate a functional form for the law $\Phi(\mathcal{A})$ before spatially averaging the rates. This situation is far more complicated and is encountered for example in case of natural photosynthesis or with photocatalytic systems using semi-conductors [20,36,44].

In all cases, the kinetic and thermodynamic formulation of a knowledge model for the photoreactor requires a sound evaluation of the mean spatial and spectral volumetric rate of radiant energy absorbed $\langle \mathcal{A} \rangle$ as explained in the next section.

**2.2. Spectral evaluation and analysis for the mean volumetric rate of radiant light energy absorbed**

For our one-dimensional purely absorbing medium with negligible reflection at the boundaries, the hemispherical input flux density $q_\cap$ is close to the flux density at the entrance of the reactor. In other words, reflectance of the slab is negligible. It is therefore possible to express the spectral mean volumetric rate of energy absorbed $\langle \mathcal{A}_\lambda \rangle$ at a given wavelength $\lambda$ as [38,43]:

$$\langle \mathcal{A}_\lambda \rangle = a_{light}\,(q_{\cap,\lambda} - q_{out,\lambda}) \qquad (4)$$

We have here introduced $a_{light}$ (equal to 36.8 m$^{-1}$ in our experimental device), the specific illuminated area of the photoreactor.

Considering that every photon whose wavelength is included in the spectral range [$\lambda_{min}$; $\lambda_{max}$] participates to the reaction enables to define the mean volumetric rate of energy absorbed (MVREA) used in the coupling model as:

$$\langle \mathcal{A} \rangle = a_{light} \int_{\lambda_{min}}^{\lambda_{max}} d\lambda\,(q_{\cap,\lambda} - q_{out,\lambda}) \qquad (5)$$

The monochromatic hemispherical photon flux density entering in front of the reactor $q_{\cap,\lambda}$ is expressed as $q_{\cap,\lambda} = q_\cap f_{\lambda,LED}$. The LED emission probability density function or the emission spectrum $f_{\lambda,LED}$ (Fig. 6, blue curve) and the mean photon flux density $q_\cap$ are accurately controlled during the experiments[36,38].

The expression of the monochromatic photon flux density at the rear $q_{out,\lambda}$ can be written as $q_{out,\lambda} = q_{out} f_{\lambda,rear}$. The mean photon flux density $q_{out}$ measured at the rear of the photoreactor is composed of transmitted blue photons generated by the LED panel ($\lambda \in$ [400 nm; 490 nm]) and photons emitted by fluorescent eosin Y ($\lambda \in$ [490 nm; 630 nm]). The probability density function at the rear $f_{\lambda,rear}$ is measured for each experiment (see material and methods section and an example is given in Fig. 6, black curve). It is thus possible to easily discriminate the blue photons and the one emitted by fluorescence.

In the following of this article, we have chosen to perform the study of $\langle \mathcal{A} \rangle$ based only on blue photons emitted by the LED panel, i.e. the fluorescence phenomenon is not considered in our current model. This choice may be justified by two arguments:

- this study is the first investigation of the radiative transfer in our photoreactor containing the reaction medium presented in the photocatalytic system section (1.1) and we restricted our analysis to spatial photonic balance only (see title) whereas fluorescence is a local radiative phenomenon;

- we are confident that perfecting our radiative model taking into account fluorescence would not change the order of magnitude of the overall quantum yield values determined hereafter.

Consequently $\langle \mathcal{A} \rangle$ is defined in equation 5 over the spectral range from 400 nm ($\lambda_{min}$) to 490nm ($\lambda_{max}$) corresponding to the blue part of the pdf $f_{\lambda,rear}$ (see shaded area in Fig. 6).

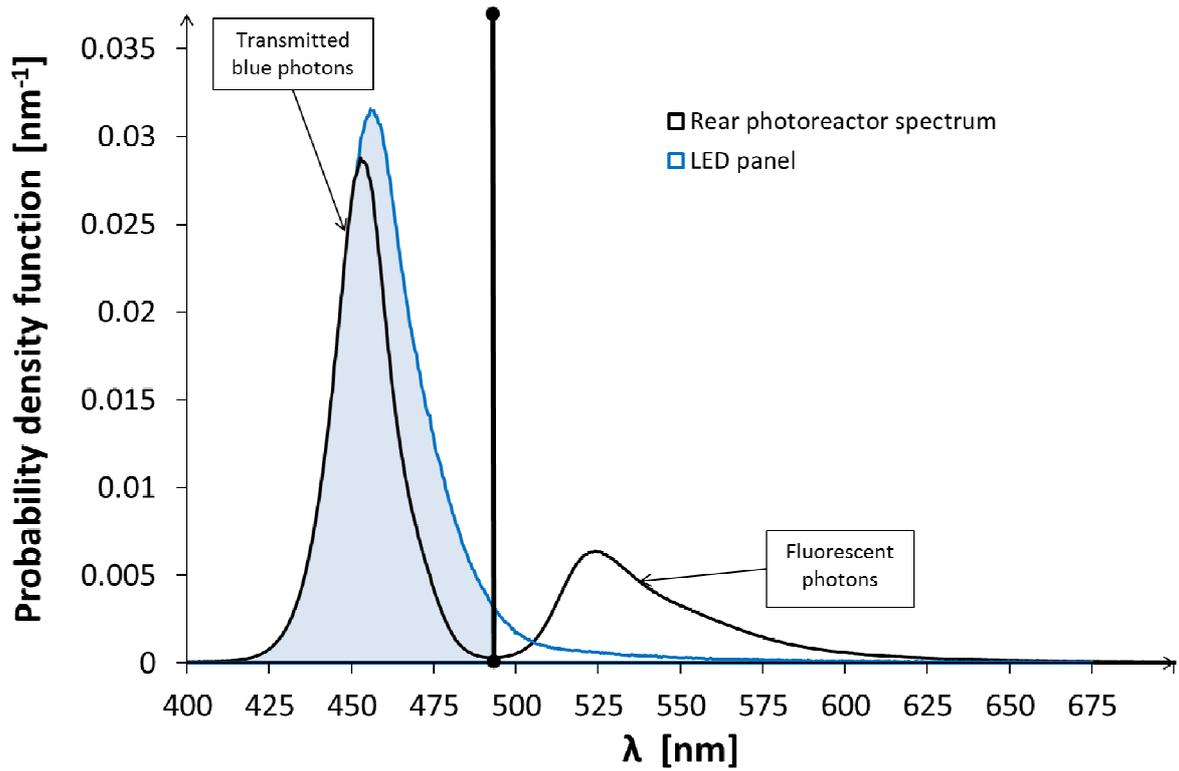

Fig. 6 – Example of the wavelength probability density function spectrum, $f_{\lambda,rear}$, of photons measured at the rear of the reactor and the LED panel emission spectrum $f_{\lambda,LED}$.

In the reaction medium, catalyst and $EY^{2-}$ are both photon absorbing species (Fig. 7) in the visible wavelength range and this point must be taken into account to go further ahead in the estimation of the mean volumetric rate of radiant light energy absorbed MVREA $\langle \mathcal{A} \rangle$. Indeed we must consider the radiant light energy only absorbed by $EY^{2-}$ because the photons absorbed by the catalyst do not lead to $H_2$ production [25]. Therefore, we introduced a factor p defined as the probability that a photon absorbed within the reactor would be absorbed by $EY^{2-}$. For a given wavelength, the probability $p_\lambda$ is defined as:

$$p_\lambda = \frac{Ea_{\lambda,EY^{2-}} C_{EY^{2-}}}{Ea_{\lambda,EY^{2-}} C_{EY^{2-}} + Ea_{\lambda,Cat} C_{Cat}} \qquad (6)$$

with $Ea_{\lambda,i}$ the absorption coefficient of i species ($EY^{2-}$ or catalyst) [m².mol⁻¹] and $C_i$ the concentration of i species ($EY^{2-}$ or catalyst) [mol.m⁻³].

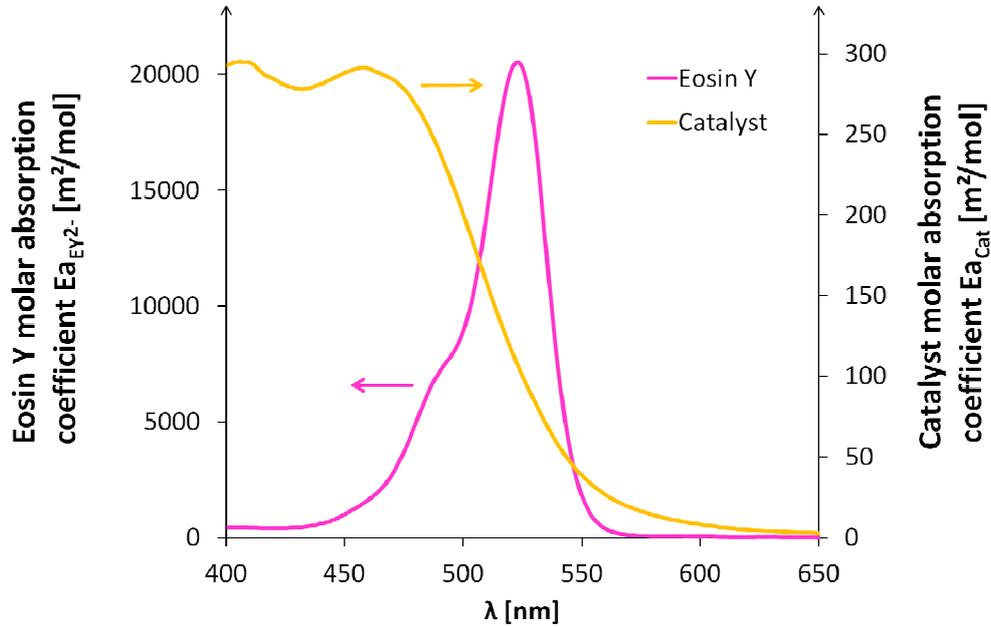

Fig. 7 – Molar absorption coefficient (Ea) of the catalyst (measured with catalyst concentration of 1.66 mM in MeOH) and of $EY^{2-}$ (measured with $EY^{2-}$ concentration of 0.024 mM in a mixture of Et₃N (10% vol.) in water at pH 10.5).

Hence, the final expression of the mean volumetric rate of radiant light energy absorbed is:

$$\langle \mathcal{A} \rangle = a_{light} \int_{\lambda_{min}=400\,nm}^{\lambda_{max}=490\,nm} d\lambda \, (q_{\cap,\lambda} - q_{out,\lambda}) p_\lambda \quad (7)$$

Every parameter necessary to evaluate numerically $\langle \mathcal{A} \rangle$ using equation 7 are known (section 1 material and methods, Fig. 6 and Fig. 7) and its values will be presented in the overall quantum yield section (0).

## 3. Hydrogen production using a model of iron-iron hydrogenase enzyme

### 3.1. Mean volumetric rate of hydrogen production

A typical experimental result for the pressure time course during $H_2$ production under irradiation of the photoreactor is presented in Fig. 8. It is clearly composed of two regimes:

- a transient regime because of unsteady hydrogen gas-liquid interphase transfer;

- a linear regime when the hydrogen transfer rate to the gas phase is equal to the reaction rate in the liquid phase enabling to apply properly the equation (1).

The temporal pressure variation $\frac{dP}{dt}$ is then extracted as the slope of the pressure time course and injected in eq. (1) using the values of the parameters previously given (the volume of gas, $V_G$, and liquid, $V_L$, and the Henry's constant for hydrogen) in order to determine accurately $\langle r_{H_2} \rangle$.

Gas chromatography analyses were carried out on the gas leaving the reactor after valve opening. The inset in Fig. 8 indicates that hydrogen was produced; traces of argon that could be present in the reactor were not detected because argon is also the gas carrier for the chromatography.

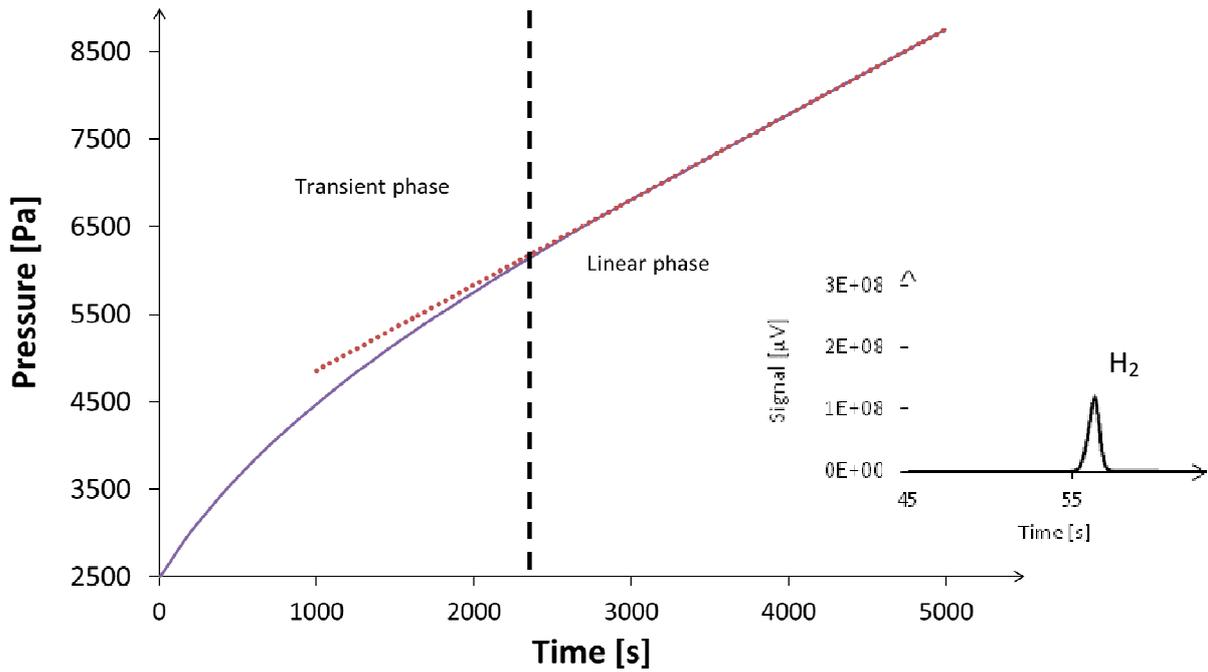

Fig. 8– Example of experimental pressure evolution presenting transient and linear regimes and gas chromatogram showing hydrogen production (inset).

Values of $\langle r_{H_2} \rangle$ were calculated for each experiment at each incident photon flux density and are displayed on Fig. 9. These results will be analyzed regarding the corresponding mean volumetric rate of radiant energy absorbed $\langle \mathcal{A} \rangle$ in the next section.

### 3.2. Overall quantum yield

At each given incident photon flux density value, $q_\cap$, corresponds an outgoing photon flux density value, $q_{out}$, and a rear spectrum $f_{\lambda,rear}$.

To determine what kind of coupling the photocatalytic system obeys, the mean volumetric rate of hydrogen production $\langle r_{H_2} \rangle$ was plotted as a function of the mean volumetric rate of radiant light energy absorbed $\langle \mathcal{A} \rangle$. Several values of incident density photon flux, $q_\cap$, were tested for each experiment at different concentration ratios of catalyst and $EY^{2-}$. To ensure the repeatability of the results and to check stability of the system, the reactor was exposed at the end of each experiment to the same photon flux density as at the beginning to verify that the pressure increase was the same. The main results of the experiments presented in Fig. 9 were obtained before a significant degradation of the photocatalytic system.

Regardless the concentration ratio of $EY^{2-}$ and catalyst, $\langle r_{H_2} \rangle$ appears proportional to $\langle \mathcal{A} \rangle$ over all the concentration range. Hydrogen production rate $\langle r_{H_2} \rangle$ is then linearly linked to the mean volumetric rate of radiant light energy absorbed $\langle \mathcal{A} \rangle$ by the overall quantum yield $\Phi$ which therefore does not depend further on $\langle \mathcal{A} \rangle$. The overall quantum yield of reaction $\Phi$ is rigorously equal to the slope of each experiment (see Fig. 9) and its experimental values are summarized in table 2.

Table 2– Summary of overall quantum yield values with their standard deviation.

| Experiments | Catalyst concentration [mM] | Eosin Y concentration [mM] | $C_{cat}/C_{EY^{2-}}$ ratio [-] | $\Phi$ values [$mol_{H_2}/mol_{h\nu}$] |
|---|---|---|---|---|
| 1 | 0.044 | 0.2 | 0.22 | $(4.7 \pm 0.2) \cdot 10^{-3}$ |
| 2 | 0.1 | 0.45 | 0.22 | $(2.2 \pm 0.2) \cdot 10^{-3}$ |
| 3 | 0.1 | 0.2 | 0.50 | $(4.4 \pm 0.4) \cdot 10^{-3}$ |
| 4 | 0.1 | 0.2 | 0.50 | $(4.6 \pm 0.4) \cdot 10^{-3}$ |
| 5 | 0.1 | 0.067 | 1.50 | $(5.7 \pm 0.3) \cdot 10^{-3}$ |
| 6 | 0.075 | 0.05 | 1.50 | $(6.1 \pm 0.9) \cdot 10^{-3}$ |

As can be seen, the overall quantum yield doesn't change when catalyst concentration varies (experiment 1,3 and 4 in Table 2) at a given $EY^{2-}$ concentration (0.2 mM); this observation has been previously shown in [21]. It is also important to note that without catalyst there is no hydrogen production.

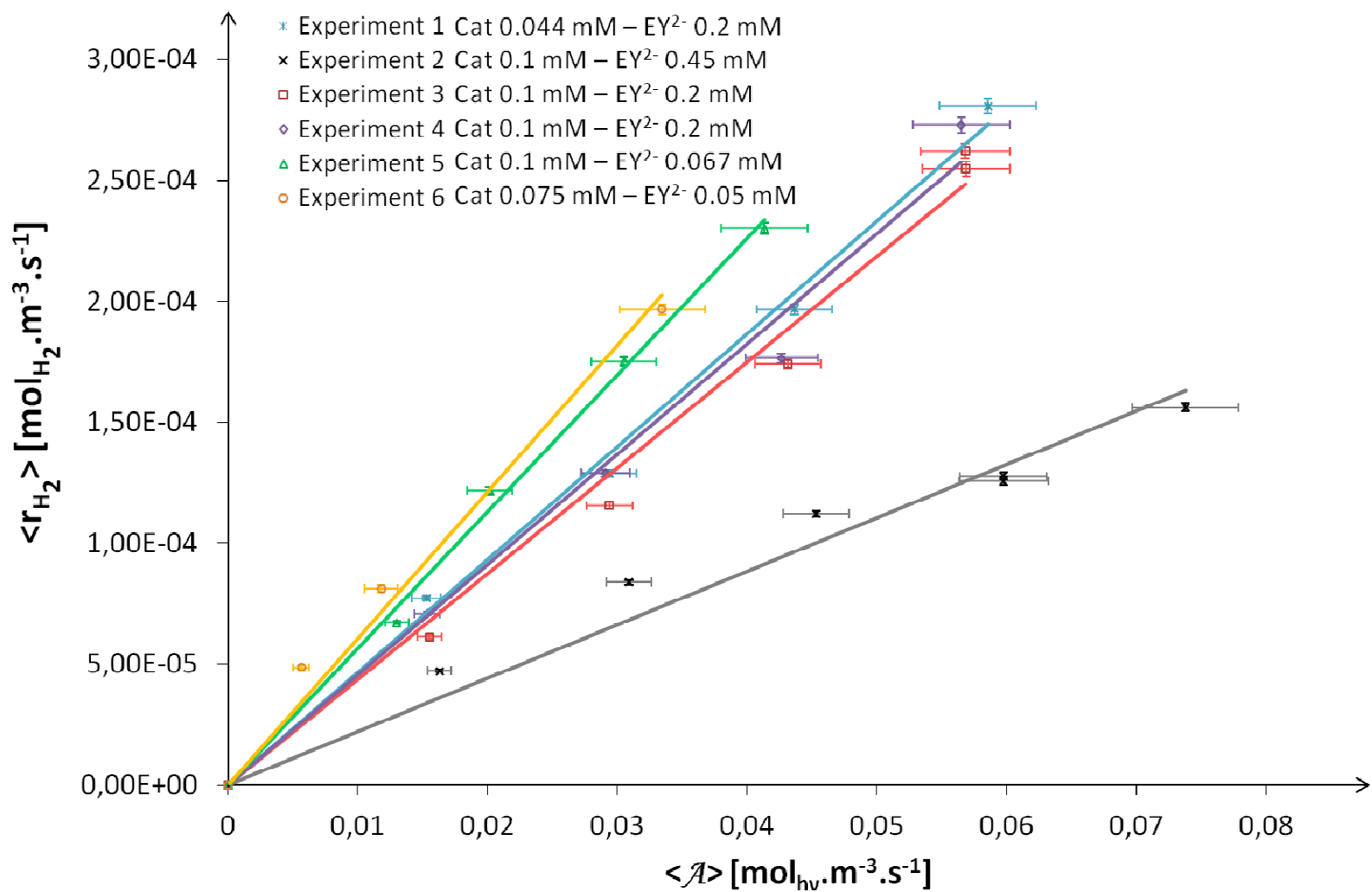

Fig. 9 – Variation of the mean volumetric rate of hydrogen production as a function of the mean volumetric rate of energy absorbed.

As also shown in Table 2, the overall quantum yield value increases when the concentration of $EY^{2-}$ decreases (considering catalyst concentration has no influence). The limited capacity (low overall quantum yield) of the photocatalytic system to convert photons into hydrogen could be explained by the low quantum yield of triplet state obtained in the chemical conditions described in this article. This could be explained by quenching phenomena[9,10,27,32,45] as presented in photocatalytic system section (1.1).

### 3.3. Specific rate of hydrogen production

From our previous experimental data, the mean molar specific rate of $H_2$ production is here defined straightforwardly as:

$$\langle J_{H_2} \rangle = \frac{\langle r_{H_2} \rangle}{C_{cat}} \qquad (8)$$

It must be noticed that this is the same quantity at the mean spatial reactor scale that the turn over frequency (TOF) generally used at molecular scale in catalytic studies. The resulting values obtained at an incident light flux density of $q_\cap = 1370\ \mu mol_{h\nu}.m^{-2}.s^{-1}$ are summarized in Table 4.

Table 4 – Summary of specific rates values with their standard deviation for each experiment for the hemispherical photon flux density $q_\cap = 1370\ \mu mol_{h\nu}.m^{-2}.s^{-1}$ entering the photoreactor.

| Experiments | Catalyst concentration [mM] | Eosin Y concentration [mM] | Specific rate $\langle J_{H_2} \rangle$ [$mol_{H_2}.mol_{cat}^{-1}.s^{-1}$] |
|---|---|---|---|
| 1 | 0.044 | 0.2 | $(4.47 \pm 0.05).10^{-3}$ |
| 2 | 0.1 | 0.45 | $(1.12 \pm 0.01).10^{-3}$ |
| 3 | 0.1 | 0.2 | $(1.74 \pm 0.02).10^{-3}$ |
| 4 | 0.1 | 0.2 | $(1.76 \pm 0.02).10^{-3}$ |
| 5 | 0.1 | 0.067 | $(1.75 \pm 0.02).10^{-3}$ |
| 6 | 0.075 | 0.05 | $(2.62 \pm 0.03).10^{-3}$ |

Except for experiment 2 with a likely quenching effect of eosin Y concentration, the specific rate for $H_2$ production at "nominal" catalyst concentration (0.1 mM) is established at $1.75 \cdot 10^{-3}$ $mol_{H_2} \cdot mol_{cat}^{-1} \cdot s^{-1}$ (Table 4). Experiments 1 and 6 show that this specific rate increases if the catalyst concentration decreases to reach a maximum value of $4.5 \cdot 10^{-3}$ $mol_{H_2} \cdot mol_{cat}^{-1} \cdot s^{-1}$ at $q_\cap = 1370$ $\mu mol_{h\nu} \cdot m^{-2} \cdot s^{-1}$ (Table 4). This confirms that mean $H_2$ production rates in the photoreactor are probably controlled by the photosystem electron transfer efficiency and that the intrinsic kinetic performances of the bio-inspired catalyst could enable far higher photoproduction rates as obtained for example with electrocatalytic systems [25].

### 3.4. Thermodynamic efficiency of the photo-reactive process

In order to compare the photocatalytic system studied here with other solar energy conversion systems, it is important to have access to the overall thermodynamic efficiency $\eta$ in the photoreactor (deduced from [17] with a good approximation):

$$\eta \cong \frac{\langle r_{H_2} \rangle \Delta G°}{\langle \mathcal{A} \rangle} \qquad (9)$$

Where $\langle r_{H_2} \rangle$ is the reaction rate [$mol_{H_2} \cdot mol_{cat}^{-1} \cdot s^{-1}$], $\langle \mathcal{A} \rangle$ the mean volumetric rate of radiant light energy absorbed [$W/m^3$], and $\Delta G° = 237$ kJ/mol is the standard free enthalpy for the hydrogen oxidation. The obtained results are summarized in table 5. Except for experiment 2 again, the values of thermodynamic efficiencies are quite close. The best value of 0.64% is reached for the lowest eosin Y and catalyst concentrations. This result, obtained in limitation by the photosystem efficiency, may appear quite low, but further consideration of the fluorescence (not taken into account in the first analysis of this paper) could decrease the actual value of $\langle \mathcal{A} \rangle$ leading to thermodynamic efficiency up to 1%. Moreover, an important characteristic of the linear coupling is that this efficiency is independent of the radiation field and then of the incident photon flux density $q_\cap$ (if optimal engineering absorption conditions are respected). It can then be obtained at an equivalent full sun ($q_\cap$ = 2000 $\mu mol_{hv}.m^{-2}.s^{-1}$) in the visible spectrum. At the opposite, the natural photosynthesis obeying to a nonlinear coupling has a thermodynamic efficiency which decreases strongly with incident flux $q_\cap$ down to 4% in a Cartesian geometry of reactor with quasi-collimated light source in the visible domain at full sun [46]. This demonstrates the superiority of the linear coupling in term of kinetic as well as energetic performances as provided by the photoreactive system studied.

Table 5 - Summary of thermodynamic efficiency values with their standard deviation (average on every hemispherical photon flux densities $q_\cap$ entering the photoreactor for each experiment).

| Experiments | Catalyst concentration [mM] | Eosin Y concentration [mM] | η values [%] |
|---|---|---|---|
| 1 | 0.044 | 0.2 | 0.42 ± 0.04 |
| 2 | 0.1 | 0.45 | 0.22 ± 0.03 |
| 3 | 0.1 | 0.2 | 0.38 ± 0.03 |
| 4 | 0.1 | 0.2 | 0.41 ± 0.04 |
| 5 | 0.1 | 0.067 | 0.51 ± 0.06 |
| 6 | 0.075 | 0.05 | 0.64 ± 0.04 |

**Conclusions and perspectives**

In this article, a bioinspired photocatalytic system, based on iron-thiolate catalyst and eosin as chromophore, has been characterized in terms of mean volumetric rate of hydrogen production $\langle r_{H_2} \rangle$. For this purpose, a high accuracy experimental bench (composed of a fully characterized light source and a gas-tight reactor) has been used. Due to its conception, it enables carrying out balances on the photonic phase and achieving a spectral radiative analysis on the photons emitted by the light source and exiting at the rear of the reactor. This balance permits estimation of the mean volumetric rate of energy absorption$\langle \mathcal{A} \rangle$. To complete this rigorous description, a p factor defined as the probability a photon absorbed within the reactor would be absorbed by eosin Y (rather than by the catalyst) has been introduced. Thanks to this analysis, it is possible to link mean spatial photon absorption and hydrogen production rates that are the key parameters in designing efficient photoprocesses for hydrogen generation. Demonstration is done that the considered bioinspired system presented in this publication belongs to photochemical system whose thermokinetic coupling is linear with a constant overall quantum yield (at fixed catalyst and dye concentrations), on the contrary to semiconductor based or biological systems.

Additionally, the bio-inspired, cheap and easily-synthesized catalyst of protons reduction has been proved to be non-limiting in the photoreactive system used. The kinetic and thermodynamic performances reached at high incident photon flux densities (equivalent to full sun) are lower than those observed in natural photosynthesis, but in the same order of magnitude. This means that, by using an engineering analysis of the photoprocess as a whole, it is possible to conclude that, probably in a near future, we will obtain sufficiently efficient and water-soluble bio-inspired molecular catalysts for protons reduction regarding photoproduction of $H_2$.

Future areas of improvement will include:

- Taking into account the fluorescence phenomena in the estimation of the mean volumetric rate of energy absorption. This point will lead to a local and more complex description of the volumetric rate of energy absorption.

- To develop a general model of the mean quantum yield that takes into account every phenomenon occurring in the bio inspired system from the photon absorption leading to singlet state to the regeneration of the bio-inspired catalyst.

- To find a more efficient dye-photosensitizer system absorbing if possible in the visible domain and with a higher electronic transfer rate to the catalyst.

**Acknowledgments**

The authors would like to thank Dr. Christophe Orain for his valuable help in implementing the chemical system in the torus reactor.

This work has been sponsored by the French government research program ''Investissements d'avenir" through the ANR program Tech'Biophyp (2011–2015), through the IMobS$^3$ Laboratory of Excellence(ANR-10-LABX-16-01) and by the European Union through theRegional Competitiveness and Employment program -2007-2013-(ERDF–Auvergne region). It is also foundedby the CNRS through thePEPS program OPTISOL_µ (2016-17). The authors acknowledge the CNRS research federationFedESol where fruitful discussions and debates take place every yearsince 2012.

**Nomenclature**

$\langle \mathcal{A}_\lambda \rangle$   spectral mean volumetric rate of radiant light energy absorbed [mol$_{hv}$.m$^{-3}$.s$^{-1}$.nm$^{-1}$]

$\langle \mathcal{A} \rangle$   mean volumetric rate of radiant light energy absorbed [mol$_{hv}$.m$^{-3}$.s$^{-1}$]

a$_{light}$   illuminated specific surface of the reactor [m$^{-1}$]

C$_i$   molar concentration of species i [mol.m$^{-3}$]

Ea$_{\lambda,i}$   molar absorption coefficient of species i at wavelength λ [m².mol$^{-1}$]

f$_{\lambda,LED}$   emission probability density function of the LED panel [nm$^{-1}$]

f$_{\lambda,REAR}$   emission probability density function at the rear of the photoreactor [nm$^{-1}$]

H$_{H_2}$(T)   Henry constant [m$^3$.Pa.mol$^{-1}$]

$\langle J_{H_2} \rangle$    mean molar specific rate of hydrogen production [$mol_{H_2} \cdot mol_{cat}^{-1} \cdot s^{-1}$]

$p_\lambda$    monochromatic probability a photon absorbed within the reactor would be absorbed by $EY^{2-}$ [-]

P    pressure [Pa]

$q_{\cap,\lambda}$    monochromatic photon flux density entering in front of the reactor [$mol_{h\nu} \cdot m^{-2} \cdot s^{-1}$]

$q_\cap$    hemispherical photon flux density entering in front of the reactor [$mol_{h\nu} \cdot m^{-2} \cdot s^{-1}$]

$q_{out,\lambda}$    monochromatic photon flux density exiting at the rear of the reactor [$mol_{h\nu} \cdot m^{-2} \cdot s^{-1}$]

$q_{out}$    photon flux density exiting at the rear of the reactor [$mol_{h\nu} \cdot m^{-2} \cdot s^{-1}$]

**r**    position in $R^3$ [-]

R    perfect gas constant [$J \cdot K^{-1} \cdot mol^{-1}$]

$\langle r_{H_2} \rangle$    mean volumetric rate of hydrogen production [$mol_{H2} \cdot m^{-3} \cdot s^{-1}$]

T    temperature [K]

t    time [s]

$V_L$    liquid volume [$m^3$]

$V_G$    headspace gas volume [$m^3$]

*Greek letter*

Φ    Overall quantum yield of the reaction [$mol_{H2} \cdot mol_{h\nu}^{-1}$]

λ    wavelength [nm]

η    thermodynamic efficiency [-]

ΔG°    standard free enthalpy of reaction [$kJ \cdot mol^{-1}$]

*Subscript*

max    maximum value

min    minimum value

out relative to the exit

**References cited in this publication**